\documentclass[
preprint,
showpacs
]{revtex4-1}
\usepackage{amsthm}
\theoremstyle{plain}
\newtheorem{theorem}{Theorem}
\newtheorem*{theorem*}{Theorem}

\newtheorem*{definition*}{Definition}

\newtheorem*{lemma*}{Lemma}

\newtheorem{remark}[theorem]{Remark}

\usepackage{braket}
\usepackage{delarray}
\usepackage{here}

\usepackage{amsmath}
\usepackage{txfonts}

\usepackage[dvipdfmx]{graphicx}
\usepackage{bm}
\usepackage{color}
\usepackage{ulem}

\usepackage{physics}

\newcommand{\ba}{\begin{array}}
\newcommand{\ea}{\end{array}}
\newcommand{\bmat}{\left(\begin{array}}
\newcommand{\emat}{\end{array}\right)}
\newcommand{\no}{\nonumber}

\newcommand{\be}{\begin{eqnarray}}
\newcommand{\ee}{\end{eqnarray}}

\begin{document}
\title{Gibbs-Bogoliubov inequality on the Nishimori line}
\author{Manaka Okuyama$^1$}
\author{Masayuki Ohzeki$^{1,2,3,4}$}
\affiliation{$^1$Graduate School of Information Sciences, Tohoku University, Sendai 980-8579, Japan}
\affiliation{$^2$International Research Frontier Initiative, Tokyo Institute of Technology, Tokyo 105-0023, Japan}
\affiliation{$^3$Department of Physics, Tokyo Institute of Technology, Tokyo 152-8551, Japan}
\affiliation{$^4$Sigma-i Co., Ltd., Tokyo 108-0075, Japan} 

\begin{abstract}  
The Gibbs-Bogoliubov inequality states that the free energy of a system is always lower than that calculated by a trial function.
In this study, we show that a counterpart of the Gibbs-Bogoliubov inequality holds on the Nishimori line for Ising spin-glass models with Gaussian randomness.
Our inequality states that the quenched free energy of a system is always lower than that calculated using a quenched trial function.
The key component of the proof is the convexity of the pressure function $\mathbb{E}\left[\log Z_{} \right]$ with respect to the parameters along the Nishimori line, which differs from the conventional convexity with respect to the inverse temperature.
When our inequality was applied to mean-field models, such as the Sherrington-Kirkpatrick model and $p$-spin model, the bound coincided with the replica-symmetric solution indicating that the equality holds.

\end{abstract}
\date{\today}
\maketitle

\section{Introduction}
The variational method is a powerful approximation technique and is used in various research areas of physics.
The Gibbs-Bogoliubov (GB) inequality~\cite{Kvasnikov,Bogoliubov,Griffiths,Isihara,Kuzemsky}, or the Gibbs-Bogoliubov-Feynman inequality~\cite{Feynman}, is one of the most famous variational inequalities in statistical physics, which shows that the quantity calculated by the variational method is always greater than or equal to the free energy of a system.
The essence of the GB inequality lies in the convexity of the pressure function $\log Z$ with respect to the inverse temperature, which immediately leads to the GB inequality~\cite{Griffiths}.

Although the GB inequality works well for ferromagnetic models, it does not perform satisfactorily for spin-glass models owing to randomness.
Thus, it leads to a natural question: Is there a variational inequality valid for spin-glass models?
In this study, we focused only on the Nishimori line and  partially answered this question.
On the Nishimori line, various exact results are obtained: exact solutions of the internal energy~\cite{Nishimori}, absence of replica symmetry breaking~\cite{NS}, and a counterpart of the Griffiths second inequalities~\cite{MNC,Kitatani}.

We prove that the counterpart of the GB inequality holds for the general Ising spin-glass model with Gaussian randomness on the Nishimori line.
The key element of the proof is the convexity peculiar to the Nishimori line.
When we control the parameters along the Nishimori line, that is, when we change the temperature and randomness simultaneously, it is possible to show that the pressure function $\mathbb{E}\qty[\log Z]$ is a convex function, which is different from the conventional convexity with respect to the inverse temperature.
Consequently, we obtain the GB inequality on the Nishimori line from convexity. 
As the ``naive counterpart" of the GB inequality in spin glass, one might think of taking the quenched average after applying the conventional GB inequality.
However, we emphasize that this ``naive counterpart" of the GB inequality is weak compared with the GB inequality on the Nishimori line obtained in the present study.

The organization of this paper is as follows.
In Sec. II, we define the model and prove the GB inequality on the Nishimori line.
In Sec. III, we apply the attained inequality to several models.
Finally, our discussion is presented in Sec. IV.


\section{Definitions and results}
We consider a generic form of the Ising spin-glass model on the Nishimori line
\be
S_A&=&\prod_{i\in A}S_i,
\\
Z_{} &=&  \Tr \qty(e^{ \sum_{A\subset\Omega} \beta_A J_{A} S_A}).
\ee
where $S_i=\pm1$, $\Omega$ is the set of sites, the sum over $A$ is over all the subsets of $\Omega$ in which interactions exist, the lattice structure adopts any form, and $\beta_A$ is the local inverse temperature of subset $A$.
The distribution of randomness of the interactions $J_A$ follows a Gaussian distribution with mean $J_{A0}$ and variance $\sigma_{A}^2$.
The thermal average $\langle \cdots \rangle$ and quenched average $\mathbb{E}\left[\cdots  \right]$ are defined by
\be
\langle f \rangle&=&\frac{\Tr\qty( f e^{ \sum_{A\subset\Omega} \beta_A J_{A} S_A})}{Z} ,
\\
\mathbb{E}\left[f  \right]
&=& \left(\prod_{A\subset\Omega} \int \frac{dJ_A}{\left(2\pi \sigma_{A}^2\right)^{1/2}} e^{-\frac{(J_{A}-J_{A0})^2}{2\sigma_{A}^2}} \right) f.
\ee
We shall be interested in the pressure function  
\be
\mathbb{E}\left[\log Z_{} \right]
&=& \left(\prod_{A\subset\Omega} \int \frac{dJ_A}{\left(2\pi \sigma_{A}^2\right)^{1/2}} e^{-\frac{(J_{A}-J_{A0})^2}{2\sigma_{A}^2}} \right) \log \Tr\left( e^{ \sum_{A\subset\Omega}\beta_A J_{A} S_A} \right).
\ee
The condition of the Nishimori line is given by
\be
\beta_{A}&=&\frac{J_{A0}}{\sigma_{A}^2 },
\ee
for all $A$.
It is convenient to introduce the parameters $x_A\ge0$~\cite{MNC,ACCM} as
\be
\beta_{A}&=&\frac{\sqrt{x_{A}}}{\sigma_{A}},
\\
J_{A0}&=&\sigma_{A} \sqrt{x_{A}},
\ee
then the pressure function depends only on $\{x_A\}$
\be
\mathbb{E}\left[\log Z_{} \right]
&=& \left(\prod_{A\subset\Omega} \int \frac{dJ_A}{\left(2\pi \right)^{1/2}} e^{-\frac{J_{A}^2}{2}} \right) \log\left( \Tr e^{  \sum_{A\subset\Omega} (\sqrt{x_A}J_{A} +x_A )S_A}\right).
\ee

Our first result is on the convexity of the pressure function, which is obtained from the calculations of previous studies~\cite{MNC,ACCM}; however, to the best of our knowledge, this has not been pointed out until now.
\begin{theorem}[convexity of Nishimori line]
Pressure function $\mathbb{E}\left[\log Z_{} \right]$ is convex for $\{x_A\}$.
\end{theorem}
\begin{proof}
For any $x_B$ and $x_C$, the derivative of the pressure function is as follows~\cite{MNC,ACCM}:
\be
\pdv{\mathbb{E}\left[\log Z_{} \right]}{x_{B}}
&=&\frac{1}{2}\mathbb{E}\qty[ 1+\langle S_B\rangle] ,
\\
\pdv{\mathbb{E}\left[\log Z_{} \right]}{x_B}{x_C}&=&\frac{1}{2}\mathbb{E}\qty[  \qty( \langle S_B S_C \rangle-\langle S_B \rangle \langle S_C \rangle)^2]. \label{sec-deri}
\ee
The following identity is useful
\be
\qty( \langle S_B S_C \rangle-\langle S_B \rangle \langle S_C \rangle)^2
&=& \left\langle \qty(S_B^1- \langle S_B\rangle) \qty(S_B^2- \langle S_B\rangle)  \qty(S_C^1- \langle S_C\rangle) \qty(S_C^2- \langle S_C\rangle) \right\rangle_{1,2}
\no\\
&=&\left\langle a_B a_C \right\rangle_{1,2} ,
\ee
where $a_B=\qty(S_B^1- \langle S_B\rangle) \qty(S_B^2- \langle S_B\rangle)$, $\langle\cdots \rangle_{1,2}$ is the thermal average with respect to the two replicas, and $S_B^1$ and $S_B^2$ are the spin in the first and second replica, respectively.
Then, we rewrite Eq. (\ref{sec-deri}) as
\be
\pdv{\mathbb{E}\left[\log Z_{} \right]}{x_B}{x_C}
&=&\frac{1}{2}\mathbb{E}\qty[  \left\langle a_B a_C \right\rangle_{1,2}] \label{sec-div-posi} .
\ee
Equation (\ref{sec-div-posi}) means that the Hessian matrix of $\mathbb{E}\left[\log Z_{} \right]$ with respect to $\{x_A\}$ is positive semidefinite, that is, $\mathbb{E}\left[\log Z_{} \right]$ is convex for $\{x_A\}$.
\end{proof}

The Gibbs-Bogoliubov inequality is a consequence of the convexity of the pressure function with respect to the inverse temperature~\cite{Griffiths}.
Using the convexity on the Nishimori line, we arrive at the Gibbs-Bogoliubov inequality on the Nishimori line.
\begin{theorem}[Gibbs-Bogoliubov inequality on Nishimori line]
For any two Ising spin-glass models on the Nishimori line,
\be
\mathbb{E}\qty[\log Z_{1}] &=& \mathbb{E}\qty[\log \Tr \qty(e^{ \sum_{A\subset\Omega_1} \beta_A J_{A} S_A})],
\\
\mathbb{E}\qty[\log Z_{0}] &=& \mathbb{E}\qty[\log \Tr \qty(e^{ \sum_{B\subset\Omega_0} \beta_B J_{B} S_B})],
\ee
the following inequality holds
\be
\mathbb{E}\left[\log Z_{1} \right]
\ge 
\mathbb{E}\left[\log Z_{0} \right]   
+  \mathbb{E}\left[\frac{1}{2}\sum_{A\subset\Omega_1} x_{A} (1+\langle S_A \rangle_0)   -\frac{1}{2}\sum_{B\subset\Omega_0} x_{B}  (1+\langle S_B \rangle_0 )\right], \label{GB-on-NL}
\ee
where  $\Omega_0$ and $\Omega_1$ are any set of sites and $\langle \rangle_0$ is the thermal average with respect to $Z_0$.
\end{theorem}
\begin{proof}
We define the interpolating pressure function on the Nishimori line as
\be
\mathbb{E}\left[\log Z_{}(t) \right]
&=& \left(\prod_{A\subset\Omega_1} \int \frac{dJ_A}{\left(2\pi \right)^{1/2}} e^{-\frac{J_{A}^2}{2}} \right) \left(\prod_{B\subset\Omega_0} \int \frac{dJ_B}{\left(2\pi \right)^{1/2}} e^{-\frac{J_{B}^2}{2}} \right)
\no\\
&& \log\left( \Tr e^{  \sum_{A\subset\Omega_1} (\sqrt{tx_A}J_{A} +tx_A )S_A + \sum_{B\subset\Omega_0} (\sqrt{(1-t)x_B}J_{B} +(1-t)x_B )S_B}\right) ,
\ee
where $\mathbb{E}\left[\log Z_{}(1) \right]=\mathbb{E}\left[\log Z_{1} \right]$ and $\mathbb{E}\left[\log Z_{}(0)\right]=\mathbb{E}\left[\log Z_{0} \right] $.
From Theorem 1, we immediately obtain 
\be
\dv[2]{\mathbb{E}\left[\log Z_{}(t) \right]}{t} \ge 0 ,
\ee
and
\be
\mathbb{E}\left[\log Z_{}(1) \right] -\mathbb{E}\left[\log Z_{}(0) \right] -\dv{\mathbb{E}\left[\log Z_{}(t) \right]}{t} \big|_{t=0}&\ge&   0. \label{divergence}
\ee
Using integration by parts, we find
\be
\dv{\mathbb{E}\left[\log Z_{}(t) \right]}{t} \big|_{t=0}&=& \frac{1}{2} \sum_{A\subset\Omega_1} x_{A}    -\frac{1}{2} \sum_{B\subset\Omega_0} x_{B}
+  \mathbb{E}\left[\frac{1}{2}\sum_{A\subset\Omega_1} x_{A} \langle S_A \rangle_0   -\frac{1}{2}\sum_{B\subset\Omega_0} x_{B}  \langle S_B \rangle_0 \right] \label{in-by-part},
\ee
where we use the identity $\mathbb{E}\left[\langle S_C \rangle_0^2\right]=\mathbb{E}\left[\langle S_C \rangle_0\right] $ for any spin product $S_C=\prod_{i\in C}S_i$ on the Nishimori line.
Combining Eqs. (\ref{divergence}) and (\ref{in-by-part}), we prove Eq. (\ref{GB-on-NL}).
\end{proof}

\begin{remark}
Note that the left-hand side of Eq. (\ref{divergence}) is called the Bregman divergence in information geometry~\cite{Amari}.
In general, for any convex function $f(t)$, the Bregman divergence is defined as $f(1)-f(0)-f'(0)\ge0$.
If we consider conventional convexity with respect to the inverse temperature, the Bregman divergence is reduced to the Kullback-Leibler divergence and we can reproduce the conventional GB inequality.
\end{remark}

\section{Examples}
In this section, we apply the GB inequality on the Nishimori line to spin-glass models and derive a mean-field approximation on the Nishimori line.
In particular, it is noteworthy that the equality holds for the Sherrington-Kirkpatrick (SK) model in the thermodynamic limit instead of the inequality.

\subsection{mean-field approximation}
First, we consider the Ising spin-glass models on the Nishimori line with the coordination number $z$
\be
\mathbb{E}\qty[\log Z_{1}] &=& \mathbb{E}\qty[\log \Tr \qty(e^{ \sum_{\langle i,j\rangle} \beta_{i,j} J_{i,j} S_iS_j})], \label{number-z}
\\
x_{i,j}&=&\beta^2.
\ee
Setting a quenched trial function as a random field 
\be
\mathbb{E}\qty[\log Z_{0}] &=& \mathbb{E}\qty[\log \Tr \qty(e^{ \sum_{i=1}^N \beta_i J_{i} S_i})],
\\
x_{i}&=&\beta^2 z q,
\ee
we apply the GB inequality on the Nishimori line  to Eq. (\ref{number-z}), which yields:
\be
\frac{1}{N}\mathbb{E}\qty[\log Z_{1}] 
&\ge& \frac{1}{N}\mathbb{E}\qty[\log Z_{0}] 
+\frac{1}{N} \frac{1}{2}\sum_{\langle i,j\rangle}x_{i,j}   -\frac{1}{2N}\sum_{i=1}^Nx_{i}  
+ \frac{1}{N} \mathbb{E}\left[\frac{1}{2}\sum_{\langle i,j\rangle}x_{i,j} \langle S_i S_j\rangle_0   -\frac{1}{2}\sum_{i=1}^Nx_{i}  \langle S_i \rangle_0\right]
\no\\
&=&\int Dy \log2\cosh\left(  \beta\sqrt{zq}y+\beta^2zq \right)
\no\\
&&+\beta^2z\qty( \frac{1}{4}     -\frac{1}{2}q
+\frac{1}{4} \qty(\int Dy \tanh\qty(  \beta \sqrt{zq}y+\beta^2 zq ))^2   -\frac{1}{2}q\int Dy \tanh\qty(  \beta \sqrt{zq}y+\beta^2 zq ) ),
\no\\
\ee
where $\int Dy=\int_{-\infty}^\infty dy e^{-y^2/2}/\sqrt{2\pi}$ and $\sum_{\langle i,j\rangle}=zN/2$.
By maximizing the right-hand side with respect to $q$, we obtain
\be
\frac{1}{N}\mathbb{E}\qty[\log Z_{1}] 
&\ge& \int Dy \log2\cosh\left(  \beta\sqrt{zq}y+\beta^2zq \right)
+\frac{z\beta^2}{4}(1-q)^2- \frac{z\beta^2q^2}{2} ,
\\
q&=&\int Dy \tanh\qty(  \beta \sqrt{zq}y+\beta^2zq ) . \label{z-sad}
\ee
This means that the GB inequality on the Nishimori line produces a mean-field approximation on the Nishimori line when a random field is chosen for a quenched trial function.

\subsection{mean-field  models}
Next, we consider the SK model on the Nishimori line
\be
\mathbb{E}\qty[\log Z_{\mathrm{SK}}] &=& \mathbb{E}\qty[\log \Tr \qty(e^{ \sum_{i,j=1}^N \beta_{i,j} J_{i,j} S_iS_j})], \label{SK-moel}
\\
x_{i,j}&=&\frac{\beta^2}{2N}.
\ee
The analytical solution in the thermodynamic limit is given by~\cite{Nishimori2,KM}
\be
\lim_{N\to\infty}\frac{1}{N}\mathbb{E}\qty[\log Z_{\mathrm{SK}}] &=&\int Dy \log2\cosh\left(  \beta\sqrt{q}y+\beta^2q \right)+\frac{\beta^2}{4}(1-q)^2- \frac{\beta^2q^2}{2} ,
\label{exact-SK}
\ee
with 
\be
q&=&\int Dy \tanh\left(  \beta\sqrt{q}y+\beta^2q \right). \label{exact-sad}
\ee

With a quenched trial function as a random field \be
\mathbb{E}\qty[\log Z_{RF}] &=& \mathbb{E}\qty[\log \Tr \qty(e^{ \sum_{i=1}^N \beta_i J_{i} S_i})],
\\
x_{i}^{RF}&=&\beta^2q,
\ee
we apply the GB inequality on the Nishimori line  to the SK model (\ref{SK-moel}).
Then Eq. (\ref{GB-on-NL}) is reduced to
\be
\frac{1}{N}\mathbb{E}\qty[\log Z_{\mathrm{SK}}] 
&\ge&\int Dy \log2\cosh\left(  \beta \sqrt{q}y+\beta^2q \right)+\frac{\beta^2}{4}-\frac{\beta^2q}{2}  
\no\\
&&+ \frac{\beta^2}{4} \qty(\int Dy \tanh\left(  \beta\sqrt{q}y+\beta^2q \right) )^2  - \frac{\beta^2q}{2} \int Dy \tanh\left(  \beta\sqrt{q}y+\beta^2q \right). \label{SK-ineq}
\ee
By maximizing the right-hand side with respect to $q$, we arrive at
\be
\frac{1}{N}\mathbb{E}\qty[\log Z_{\mathrm{SK}}]&\ge&\int Dy \log2\cosh\left(  \beta\sqrt{q}y+\beta^2q \right)+\frac{\beta^2}{4}(1-q)^2- \frac{\beta^2q^2}{2} , \label{replica-bound}
\ee
with 
\be
q&=&\int Dy\tanh\left(  \beta\sqrt{q}y+\beta^2q \right),
\ee
which coincides with the exact solution of Eqs. (\ref{exact-SK}) and (\ref{exact-sad}).
Therefore, the equality of the GB inequality on the Nishimori line holds for the SK model in the thermodynamic limit.

Similar results also hold for the $p$-spin glass model with a random field on the Nishimori line,
\be
\mathbb{E}[\log Z_p]&=& \mathbb{E}\qty[\log \Tr(\exp( \sum_{i_1<\cdots <i_p}\beta_{i_1,\cdots,i_p}J_{i_1,\cdots,i_p}\sigma_{i_1}\cdots\sigma_{i_p}+\sum_i \beta_i h_i \sigma_i ))],
\\
x_{i_1,\cdots, i_p}&=&\frac{\beta^2p!}{2N^{p-1}} ,
\\
x_i&=& \beta^2 h,
\ee
where $h\ge0$ and $p$ is any positive integer.
By choosing a quenched trial function as a random field with the parameter $x_{i}^{RF}=p\beta^2q^{p-1}/2+\beta^2h$ and maximizing the GB inequality on the Nishimori line with respect to $q$, 
we arrive at 
\be
&&\frac{1}{N}\mathbb{E}[\log Z_p] 
\no\\
&\ge&
\int Dz \log2\cosh\left(   z\sqrt{\frac{p}{2}\beta^2q^{p-1}+\beta^2 h}+\frac{p}{2}\beta^2q^{p-1} +\beta^2 h\right)
+\frac{\beta^2}{4} (1-pq^{p-1})
+ \frac{\beta^2}{4} (1-p) q^p, \label{p-spin-ineq}
\ee
with
\be
q&=&\int Dz \tanh\left(  z\sqrt{\frac{p}{2}\beta^2q^{p-1}+\beta^2 h}+\frac{p}{2}\beta^2q^{p-1}+\beta^2 h\right) .
\ee
We note that the right-hand side of Eq. (\ref{p-spin-ineq}) coincides with the exact solution in the thermodynamic limit~\cite{Nishimori2,KM,BM2}.

Finally, while the inequalities (\ref{SK-ineq}) and (\ref{p-spin-ineq}) for mean-field models have already been obtained in Ref. \cite{KM,BM2}, our derivation based on the GB inequality on the Nishimori line is much simpler.
Additionally, in  the case of odd $p$, our inequality (\ref{p-spin-ineq}) is true for all $h\ge0$, whereas previous studies~\cite{KM,BM2} were limited to Lebesgue almost every $h\ge0$.
Thus, our derivation is slightly better than previous studies.

\section{Discussions}
We showed that the counterpart of the GB inequality holds on the Nishimori line for general spin-glass models with Gaussian randomness.
The convexity on the Nishimori line plays an essential role in the derivation.

When a random field is chosen for a quenched trial function, the GB inequality on the Nishimori line recovers the mean-field approximation on the Nishimori line. 
Moreover, the equality of the GB inequality on the Nishimori line holds for the Sherrington-Kirkpatrick model in the thermodynamic limit, which corresponds to the fact that the equality of the conventional GB inequality holds for the Curie-Weiss model in the thermodynamic limit when we set a trial function as the magnetic field.
These results show that the GB inequality on the Nishimori line perfectly corresponds to the conventional GB inequality.


When the GB inequality on the Nishimori line is applied to $p$-spin model with odd $p$, the obtained inequality $(\ref{p-spin-ineq})$ is valid for any $h\ge0$ whereas the corresponding inequalities in previous studies~\cite{KM,BM2} are limited to Lebesgue almost every $h\ge0$.  
Therefore, our inequality is slightly stronger than previous studies~\cite{KM,BM2}.
It is expected that the GB inequality on the Nishimori line can also be applied to more complicated mean-field models~\cite{ACCM,ACCM2,BM,BM2}.

While we only considered Gaussian randomness, the Nishimori line exists in other randomness, such as $\pm J$ Ising model. It is interesting to determine  whether the GB inequality on the Nishimori line holds for other randomness.

It is important to see how the pressure function behaves in the case of $J_{A0}=0$, which is most interesting to us.
Recent studies~\cite{AC,AC2} have proven that when the inverse temperature scale is changed to $\sqrt{\beta}$, $\mathbb{E}\qty[\log \Tr \qty(e^{ \sqrt{\beta}\sum_{A\subset\Omega}  J_{A} S_A})]$ is concave with respect to $\beta$ in several mean-field spin-glass models using exact solutions of the thermodynamic limit.
There is no guarantee that this property holds for general spin-glass models, and it would be interesting to investigate its behavior in finite-dimensional models.


This work was financially supported by JSPS KAKENHI Grant Nos. 19H01095,  20H02168, and 21K13848.



\begin{thebibliography}{99}

\bibitem{Kvasnikov}
I. A. Kvasnikov, 
Dokl. Akad. Nauk SSSR \textbf{110}, 755 (1956).

\bibitem{Bogoliubov}
N. N. Bogoliubov, 
Dokl. Akad. Nauk USSR \textbf{119}, 244 (1958)

\bibitem{Griffiths}
R. B. Griffiths,  
J. M. Phys. \textbf{5}. 1215 (1964)

\bibitem{Isihara}
A. Isihara, 
J. Phys. A: General Phys. \textbf{1}, 539 (1968).


\bibitem{Kuzemsky}
A. L. Kuzemsky, 
Int. J. Mod. Phys. B \textbf{29}, 1530010 (2015).

\bibitem{Feynman}
R. P. Feynman,
Phys. Rev. \textbf{97}, 660 (1955).


\bibitem{Nishimori} 
H. Nishimori, 
Prog. Theor. Phys. \textbf{66}, 1169 (1981).

\bibitem{NS}
H. Nishimori and D. Sherrington,
In: AIP Conference Proceedings, vol. 553, p. 67 (2001).

\bibitem{MNC}
S. Morita, H. Nishimori, and P. Contucci,
J. Phys. A \textbf{37}, L203 (2004).

\bibitem{Kitatani}
H. Kitatani,
J. Phys. Soc. Jpn. \textbf{78}, 044714 (2009).

\bibitem{ACCM}
D. Alberici, F. Camilli, P. Contucci, and E. Mingione,
Comm. Math. Phys.  \textbf{387}, 1191 (2021).

\bibitem{Amari}
S-i. Amari,
\textit{Information geometry and its applications.}
(Springer, 2016).


\bibitem{Nishimori2}
H. Nishimori,
\textit{Statistical Physics of Spin Glasses and Information Processing: An Introduction}
 (Oxford University Press, Oxford, 2001).
 

\bibitem{KM}
S. B. Korada and N. Macris,
J. Stat. Phys. \textbf{136}, 205 (2009).
\bibitem{BM}
J. Barbier and N. Macris,
J. Phys. A \textbf{52}, 294002 (2019).
\bibitem{BM2}
J. Barbier and N. Macris,
Probab. Theory Relat. Fields \textbf{174}, 1133 (2019).
\bibitem{ACCM2}
D. Alberici, F. Camilli, P. Contucci, and E. Mingione,
J. Stat. Phys. \textbf{182}, 2 (2021).



\bibitem{AC}
A. Auffinger and W.-K. Chen,
Comm. Math. Phys.  \textbf{348}, 751 (2016).

\bibitem{AC2}
A. Auffinger and W.-K. Chen,
Electron. J. Probab. \textbf{22}, 1 (2017).

\end{thebibliography}
\end{document}